\documentclass[11pt,a4paper]{article}

\usepackage[utf8]{inputenc}
\usepackage[T1]{fontenc}
\usepackage{mathptmx}  
\usepackage[margin=2.5cm]{geometry}
\usepackage{graphicx}
\usepackage{amsmath}
\usepackage{amssymb}
\usepackage{algorithm}
\usepackage{algpseudocode}
\usepackage[version=4]{mhchem}
\usepackage{booktabs}
\usepackage{multirow}
\usepackage{hyperref}
\usepackage{url}
\usepackage{float}
\usepackage{natbib}  
\usepackage{setspace}
\usepackage{authblk}  

\title{\textbf{Predictive control of blast furnace temperature in steelmaking with hybrid depth-infused quantum neural networks}}

\author[1]{Nayoung Lee}
\author[2]{Minsoo Shin}
\author[3]{Asel Sagingalieva}
\author[3]{Arsenii Senokosov}
\author[3]{Matvei Anoshin}
\author[3]{Ayush Joshi Tripathi}
\author[3]{Karan Pinto}
\author[3]{Alexey Melnikov\thanks{Corresponding author: alexey@melnikov.info}}

\affil[1]{POSCO Holdings Inc., 6261, Donghaean-ro, Nam-gu, Pohang-si, Gyeongsangbuk-do, Republic of Korea}
\affil[2]{POSCO, 6261, Donghaean-ro, Nam-gu, Pohang-si, Gyeongsangbuk-do, Republic of Korea}
\affil[3]{Terra Quantum AG, Kornhausstrasse 25, 9000 St.~Gallen, Switzerland}

\date{}

\begin{document}

\maketitle

\begin{abstract}
\noindent
Accurate prediction and stabilization of blast furnace temperatures are crucial for optimizing the efficiency and productivity of steel production. Traditional methods often struggle with the complex and non-linear nature of the temperature fluctuations within blast furnaces. This paper proposes a novel approach that combines hybrid quantum machine learning with pulverized coal injection control to address these challenges. By integrating classical machine learning techniques with quantum computing algorithms, we aim to enhance predictive accuracy and achieve more stable temperature control. For this we utilized a unique prediction-based optimization method. Our method leverages quantum-enhanced feature space exploration and the robustness of classical regression models to forecast temperature variations and optimize pulverized coal injection values. Our results demonstrate a significant improvement in prediction accuracy over 25\% and our solution improved temperature stability to $\pm 7.6\,^\circ\mathrm{C}$ of target range from the earlier variance of $\pm 50\,^\circ\mathrm{C}$, highlighting the potential of hybrid quantum machine learning models in industrial steel production applications. This work demonstrates the practical viability of quantum-enhanced AI for real-world manufacturing control, providing a framework that can be adopted by steel producers seeking to improve process efficiency.

\vspace{0.5cm}
\end{abstract}

\section{Introduction}
\label{sec:intro}

Steel forms the backbone of the global economy, primarily due to its strength, durability, and versatility. Steel manufacturing is a highly complex and energy intensive endeavor. Steel production is a cornerstone of modern industry, and the blast furnace is at the heart of this process \citep{Carpenter2006, Li2019}. In a blast furnace, raw materials such as iron ore, coke, and limestone are subjected to high temperatures to produce molten iron, which is then converted into steel \citep{Jiang2023, Zhang2023, Calix2023}. Maintaining optimal temperature within these furnaces is critical for maximizing efficiency, reducing energy consumption, and ensuring high-quality output. The dynamic and highly non-linear nature of temperature variations within blast furnaces poses significant challenges to achieving stable and precise control \citep{Xu2024}.

One of the strategies employed to enhance the performance of blast furnaces is Pulverized Coal Injection (PCI), which involves injecting fine coal particles into the blast furnace to partially replace coke as a fuel and reducing agent. PCI offers numerous advantages, including cost reduction and improved control over the combustion process. However, effectively managing the injection rate and distribution of pulverized coal to maintain stable furnace temperatures requires sophisticated control strategies \citep{Yuan2015}.

Machine learning models have demonstrated high performance in solving complex predictive problems in a wide range of domains \citep{Krenn2023}, from economics \citep{Stock2006, Paquet2022, Kea2024} to energy \citep{Feng2025} and industry \citep{Sebastianelli2021, Kim2019}. In particular, recurrent neural networks like long short-term memory (LSTM) networks \citep{Hochreiter1997, Gers2000} have been successful in modeling complex temporal patterns, including non-linear interactions, long-term trends, and cyclical fluctuations. Despite these advances, the performance of purely classical machine learning models may be constrained by factors such as limited or noisy datasets, high-dimensional inputs, and intricate fault dynamics \citep{Bishop2007, Goodfellow2016}.

Recently, quantum computing has begun to transition from theoretical prototypes to early practical implementations, leveraging fundamental principles like entanglement and superposition to process information in ways that are difficult or even impossible for classical computers to mimic \citep{Nielsen2010, Biamonte2017, Melnikov2023}. This paradigm shift has spurred the development of quantum algorithms aimed at boosting performance in domains such as optimization, simulation, and cryptography, laying the groundwork for more specialized applications \citep{Montanaro2016, Preskill2018, Majumder2024}. Within this emerging ecosystem, quantum machine learning (QML) has attracted considerable interest for its potential to handle complex problems that suffer from data scarcity or high-dimensional feature spaces \citep{Rebentrost2014, Ciliberto2018, Schuld2018, Cao2017, Kerenidis2019a, Kerenidis2019b, Dawid2022, Flamini2024}. QML models exploit high-dimensional Hilbert spaces to embed input features, enabling the representation of intricate correlations with fewer parameters compared to purely classical methods \citep{Havlicek2019, Schuld2019, Sedykh2024, Senokosov2024}. This ability to map data into richer feature representations can offer significant advantages when dealing with nonstationary or noisy signals, as commonly encountered in industrial maintenance and prognostic settings \citep{Kurkin2025, Sagingalieva2025a, Emmanoulopoulos2022}.

A practical way to harnessing these quantum resources is through hybrid quantum-classical neural networks (HQNNs), which seamlessly combine both classical and quantum layers in a unified architecture \citep{Havlicek2019, Bischof2025, Sun2025}. In such models, quantum circuits often act as specialized components for data transformation or feature encoding, while the broader network structure encompassing backpropagation, parameter updates, and other large-scale operations remains anchored in conventional machine learning frameworks. This hybrid scheme exploits quantum effects in targeted segments of the computational pipeline, without sacrificing the scalability and reliability of well-established classical methods \citep{Rainjonneau2023, Broughton2020, Haboury2023, Landman2022, Sagingalieva2023b, Sedykh2025, Laskaris2026}. Early research suggests that HQNNs can achieve competitive or even superior performance compared with purely classical deep learning techniques, frequently demonstrating enhanced robustness against overfitting \citep{Abbas2021, Berberich2024}.

Our methodology merges machine learning with quantum algorithms to form a hybrid system for real-time temperature forecasting and PCI adjustment. By enhancing predictive accuracy and control stability, this HQNN-based approach \citep{Benedetti2019, Abbas2021, Skolik2021} promises the potential to significantly improve the efficiency and productivity of steel production.

The remainder of this paper is organized as follows. Section~\ref{sec:background} provides background on steel making in blast furnaces and existing temperature prediction approaches. Section~\ref{sec:methods} describes our materials and methods, including the industrial dataset, feature engineering, and model architectures. Section~\ref{sec:results} presents our experimental results. Section~\ref{sec:discussion} discusses the implications and limitations of our approach. Finally, Section~\ref{sec:conclusion} concludes the paper.

\section{Background and Related Work}
\label{sec:background}

\subsection{Steel Making in Blast Furnace}

\subsubsection{Process Overview}

One of the key processes for producing iron is melting iron ore and through a reduction process, creating liquid iron \citep{Geerdes2020}. This process of making hot metal is called the blast furnace process. The main reactions are as follows:

(1) Coke combustion: Coke (Carbon) and \ce{O_2} are oxidized at high temperatures to produce \ce{CO} gas:
\begin{equation}
\ce{C + O_2 -> CO}
\end{equation}

(2) Iron ore reduction: \ce{CO} gas separates oxygen from iron ore to produce pure Fe:
\begin{align}
\ce{3Fe_2O_3 + CO &-> 2Fe_3O_4 + CO_2}\\
\ce{FeO + CO &-> Fe + CO_2}
\end{align}

In the upper part of the furnace:
\begin{equation}
\ce{FeO + CO -> Fe +CO_2}
\end{equation}

In the lower part:
\begin{equation}
\ce{FeO + C -> Fe + CO}
\end{equation}

The blast furnace process consists of raw material handling equipment, charging equipment, blast furnace main body, hot blast stove/air blower equipment, Pulverized Coal Injection equipment, burden and raw materials equipment, and gas cleaning equipment, and operates continuously 24 hours a day without stopping. The primary equipment in operation is the blast furnace main body, divided into five sections vertically: the gas riser and rotating chute, the throat, the lumpy zone, the cohesive zone where burden materials are dissolved and volume shrinks at the bosh, and the bottom where the wind tuyere is located for fuel combustion and molten hot metal is stored at the hearth with the tap hole exit \citep{Agarwal2010}. The point where the pulverized coal injection occurs is at the bottom of the blast furnace hearth. The tuyere applies heat to the mixture inside at a temperature of 1200°C and 4.0 bars.

PCI reduces the cost of raw material by substituting some of the high-cost coke with pulverized coal and medium oil injected into the blower at the bottom of the blast furnace \citep{Carpenter2006, Zhou2016}. This involves inserting an injection lance into the internal duct of the tuyere at the bottom of the blast furnace for blowing.

\subsubsection{Temperature Prediction Challenges in Blast Furnaces}

\begin{figure}[htbp]
    \centering
    \includegraphics[width=\textwidth]{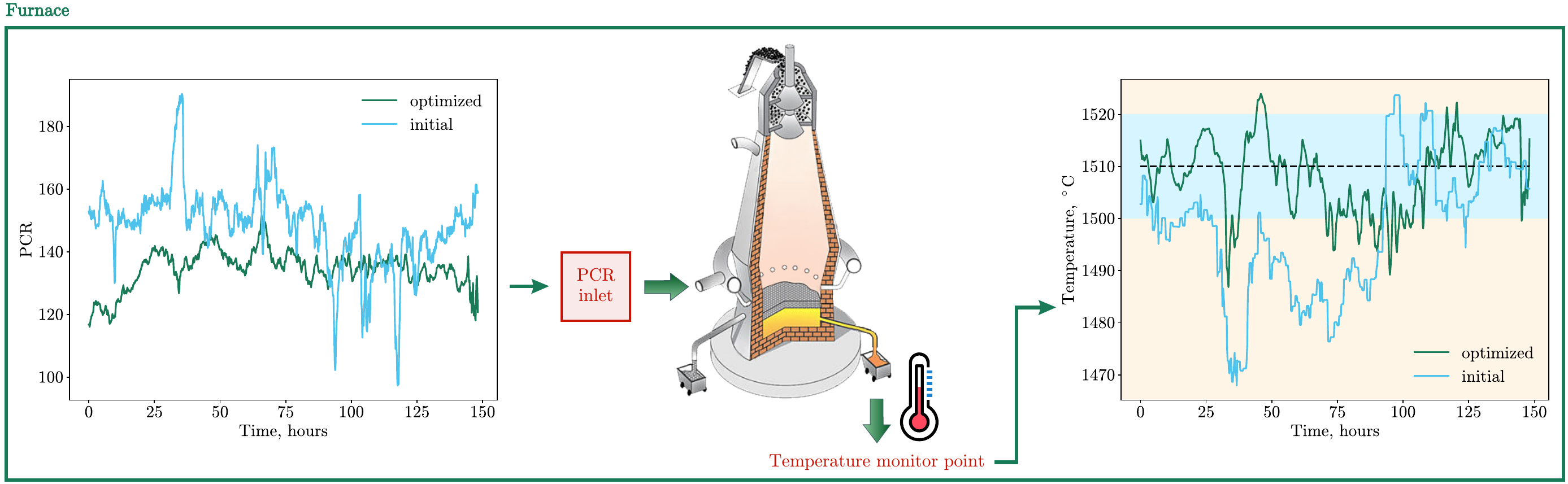}
    \caption{The left plot shows the original (initial) vs. optimized PCI injection rates over time. The right plot presents the resulting temperature trends. The optimized PCI policy (green) successfully maintains the temperature within the target operational window (blue-shaded band between 1500$^\circ$C and 1510$^\circ$C), while the initial policy (blue) leads to significant deviations and instability.}
    \label{fig:optim}
\end{figure}

In the iron making process, it is important to produce molten iron at a consistent temperature. Smooth iron making leads to stable subsequent processes. However, controlling the temperature consistently in the blast furnace is difficult due to factors such as the inability to see inside, limited control points, and time delays between charging, control, and tapping \citep{Zhou2015, Zhang2019a, Bhattacharya2005}.

The challenge of controlling the temperature originates from the delay between the control action and its effect, which cannot be clearly defined due to uncertainties in the relationship between them. Therefore, achieving stable temperature control requires understanding the relationship between the current control action and future temperature changes, which is not easy to predict accurately. To address this sub task, a sensor system that can provide real-time information and analysis of temporal relationships among these sensors is needed, along with a model that can accurately predict the temperature.

Time delay characteristics are particularly problematic in blast furnace operations. The blast furnace is large and inaccessible, making it impossible to track the materials inside or measure the high-temperature environment accurately. The lack of sensors that can withstand the extreme conditions of the blast furnace contributes to the inability to collect internal data. Time delays result from differences in transmission time, reaction time, and spatial-temporal distributions within the smelting equipment. According to field experts, the time delay between PCI charging and molten iron temperature is 2--3 hours.

If the temperature drops too low and the iron ore fails to melt properly, there is a risk of blockages in the upper portion of the blast furnace. Conversely, if the temperature drops around the relatively lower tuyeres, the molten iron might solidify again, potentially causing the dangerous phenomenon known as ``back lamination.'' In severe cases, explosions called ``hang-ups'' can occur, jeopardizing not only production but also the safety of workers near the blast furnace.

Even considering only PCI operation, approximately 110 tons/hr of coal for one blast furnace and 70 tons/hr for other furnaces are injected (at current costs of 30--60 thousand KRW per ton). Operators typically aim to keep the furnace within $\pm 15\,^\circ\mathrm{C}$ of the target by setting a higher target temperature as a buffer. By stabilizing the hearth, i.e., controlling the molten iron temperature more precisely, the target temperature setpoint can be lowered, significantly reducing overall fuel usage.

\subsection{Temperature Prediction and Control Strategies}

Temperature prediction and control for the blast furnace has quite a long history. Prediction methods are divided into mathematical model-based methodology \citep{Chu2006} and data-driven methodology. Sometimes the silicon content replaces the temperature since it is a good indicator and measuring the temperature of the molten iron is dangerous. Predicting the silicon content is also divided into mathematical modeling \citep{Saxen2007, Nurkkala2011} and data-driven methodologies \citep{Qiu2009, Saxen2012, Zhang2018, Chen2019, Jiang2022}.

To control the heat of a blast furnace, methods such as regulating the amount of input fuel and adjusting the air supply are available. Among these, for the primary purpose of stabilizing the furnace heat during sudden fluctuations, the control of the PCI rate is given priority. PCI consists of fine coal particles and is injected through the tuyeres.

The rate of pulverized coal injection is defined as:
\begin{equation}
\text{PCI}\;(\text{ton}/\text{hr}) = R_c - R_d \cdot P \cdot \frac{1000}{24},
\end{equation}
where $R_c$ is corrected reductant rate, $R_d$ is dropped coal rate and $P$ is hot metal production from the blast furnace.

The amount of pulverized coal injection is calculated as part of the total reductant ratio (RAR) which is the sum of real time PCR and real time CR:
\begin{equation}
\text{RAR} = \left( \frac{\text{PCI}_\text{total}}{P_\text{real}} \cdot 24 + \frac{C_c}{C_{pb}} \right) \cdot 1000,
\end{equation}
where $\text{PCI}_\text{total}$ is total PCI rate, $P_\text{real}$ is real time production, $C_c$ is total coke house spt calculation, $C_{pb}$ is pig production burden change. The optimal PCI rate is determined by how stably the furnace heat can be controlled. Thus, in order to decide the appropriate amount of coal to inject in the future, it is essential to predict how the furnace temperature will behave in the near future under different PCI scenarios.

In our approach, feature reduction techniques are employed to utilize a broad array of sensor data and identify which signals have the strongest influence on the molten iron temperature. By doing so, we aim to base the PCI control on a more comprehensive representation of the furnace state.

\section{Materials and Methods}
\label{sec:methods}

\subsection{Industrial Dataset and Measurement Setup}

Data was obtained from a blast furnace at POSCO's Gwangyang Steel Works, specifically from the year 2023. Data has been recorded every minute and includes readings from approximately 580 sensors. These sensors measure variables such as tuyere velocity, \ce{CO} and \ce{H_2} gas utilization, stave temperature by position, hearth wall temperature by position, hot metal tapped temperature, cast speed, furnace top pressure, furnace gas temperature, stack pressure, air humidity, as well as the composition of pig iron and slag.

The presented data consist of two parts, each covering a time span of one month. These datasets contain 580 sensor columns, including PCI, as well as a timestamp column and four columns corresponding to temperature values at different tap holes of the furnace. The average of these four columns is considered as the temperature.

For tackling the temperature time-series prediction and the subsequent optimization task, we designed a feature engineering workflow with three main steps:
\begin{enumerate}
    \item Data Selection, Treatment and Preprocessing
    \item Data Dimensionality Reduction 
    \item Data Discretization
\end{enumerate}

\subsubsection{Temperature Measurement and Cleaning}

To calculate the optimal PCI injection amount, the currently observed factors include daily iron production, heat load, \ce{CO} utilization, \ce{H_2} utilization, \ce{Si} amount in molten iron, and molten iron temperature. Daily iron production is the production volume calculated based on the amount of input fuel materials, and heat load is the load calculated using the cooling water flow rate, inlet temperature, and outlet temperature. \ce{CO} utilization is the ratio calculated using the amounts of \ce{CO} and \ce{CO_2} gases analyzed from the exhaust gas, while \ce{Si} amount in molten iron is a value obtained through sampling and analysis of the hot metal in the laboratory.

In this study, the molten iron temperature, frequently used at POSCO, was selected as the target. The measurement is performed at a point approximately 2 blocks away from the tap hole where the molten iron flows out of the furnace and passes through the runner. There are a total of 4 tap holes located in the cardinal directions of the circle, and the measurement interval starts about 100--120 minutes after the tap hole is opened, measuring once per hour with respect to the runner temperature on that side.

As with typical outlier handling, missing values represented as 0 or null are handled, manual measurement error correction is performed, and an additional Inter Quartile Range (IQR)-based error correction is applied as shown in Table~\ref{tab:iqr}. In the case of IQR-based outlier correction, a distance of 5 IQR instead of the typical 1.5 IQR distance between the 1st and 3rd quartiles is set, considering the data characteristics of the blast furnace operation.

\begin{table}[htbp]
\centering
\caption{IQR calculation for preprocessing of Molten Iron Temperature Measurement Data. For calculating the IQR, the minimum, median, maximum, Q1, and Q3 values were obtained for each sensor. The standard 1.5 IQR did not match the anomaly range of the blast furnace, and thus it was adjusted to 5 IQR for anomaly value processing.}
\label{tab:iqr}
\begin{tabular}{lcccc}
\toprule
Feature & Temp 1 & Temp 2 & Temp 3 & Temp 4\\
\midrule
Sample Size & 47520 & 47520 & 47520 & 47520\\
Minimum & 1301.832 & 1267.96 & 1208.67 & 1321.189\\
Q1 & 1500 & 1503.24 & 1502.87 & 1500\\
Median & 1509.14 & 1521.06 & 1505.6 & 1510.83\\
Q3 & 1523.52 & 1525.05 & 1520.107 & 1521.98\\
Maximum & 1589.26 & 1556.671 & 1556.854 & 1560.45\\
IQR ($=Q3-Q1$) & 23.52 & 21.81 & 17.24 & 21.98\\
$Q1 - 5\times \text{IQR}$ & 1382.4 & 1385.64 & 1385.27 & 1382.4\\
$Q3 + 5\times \text{IQR}$ & 1617.6 & 1620.84 & 1620.47 & 1617.6\\
\bottomrule
\end{tabular}
\end{table}

\subsubsection{Feature Selection and Dimensionality Reduction}

Due to the extreme environment of the blast furnace, it is not possible to directly measure internal states, like internal temperature or pressure profiles, except at a few boundary points. Measurements are available only from the outside of the furnace (e.g., wall temperatures) or from outputs, like the tapped molten iron. This means control is largely reactive: one can only respond to changes after they have affected measured outputs.

Our goal was to develop a predictive algorithm that could be practically implemented in a real plant control system. We employed a gradient boosting (GB) model to evaluate feature importance with respect to two prediction targets: the furnace temperature and the PCI rate. We trained GB models (using decision-tree ensembles) on the dataset for two separate tasks: one model was trained to predict the temperature using all sensor inputs, and another model to predict the PCI rate. By examining the learned importance weights in each model's decision trees, we obtained an importance score for each input feature.

Using these importance scores, we selected the top $N$ features as our reduced feature set. Ultimately, we settled on 25 features in total (combining the top 19 from the temperature-targeted GB and the top 6 from the PCI-targeted GB). The list of the top 26 features by importance is given in Table~\ref{tab:features}, along with their linear correlation with the temperature.

\begin{table}[htbp]
\centering
\caption{Influence (feature importance) on temperature using the GB for top 26 features. The order is ascending by influence. Corr is correlation on temperature.}
\label{tab:features}
\begin{tabular}{lcc}
\toprule
No. & Influence & Corr\\
\midrule
1 & 15.56 & -0.67\\
2 & 11.32 & -0.58\\
3 & 7.66 & -0.44\\
4 & 7.32 & -0.11\\
5 & 5.88 & -0.43\\
6 & 5.64 & -0.26\\
7 & 4.63 & 0.45\\
8 & 4.28 & -0.46\\
9 & 3.81 & -0.38\\
10 & 3.71 & -0.18\\
\bottomrule
\end{tabular}
\end{table}

\subsubsection{Temporal Discretization of Time Series}

Our ultimate goal is to establish a relationship between changes of PCI and subsequent changes in temperature, suitable for optimization. To do this, we considered time-window averaging to reduce noise and align the data to consistent time steps.

If we denote by $T^{(t)}$ the temperature at time $t$, we define the discretized temperature over a window of length $l_{\text{window}}$ as a moving average:
\begin{equation}
T^{(t_0)}_{l_{\text{window}}} = \frac{1}{l_{\text{window}}}\sum_{t = t_0}^{t_0 +l_{\text{window}}} T^{(t)}
\label{eq:discr}
\end{equation}
and then we down-sample (shift the window by $l_{\text{window}}$ for the next data point). We applied this to both the temperature and the input features to create a discretized dataset for model training.

\subsection{Temperature Prediction Models}

In order to determine the optimal PCI values that achieve the required temperature for steel production, we first need a model that predicts how the furnace temperature will respond to changes in PCI. This is essentially a temperature forecasting problem given time-series data.

To solve this prediction task, we developed a specialized machine learning model $\text{M}_\text{T}$. The core of this model is an LSTM \citep{Hochreiter1997, Tian2019} network, which consists of gates implementing the memory effect. In our configuration, we use an LSTM layer with a hidden state size of 143. This LSTM takes as input a sequence of 24 time steps of discretized sensor data (with 27 features per time step). Thus, the LSTM processes an input of shape $24 \times 27$, which corresponds to the most recent 24 discrete time units (e.g., $24 \times 10$ minutes = 240 minutes of data). The LSTM's output is then fed into a fully connected layer that maps the hidden state to a future temperature vector of length 5, predicting 5 future discrete time points. We train this model by backpropagating the error L2 loss function between the predicted temperature vector and the actual future temperature vector.

\begin{figure}[htbp]
    \centering
    \includegraphics[width=\textwidth]{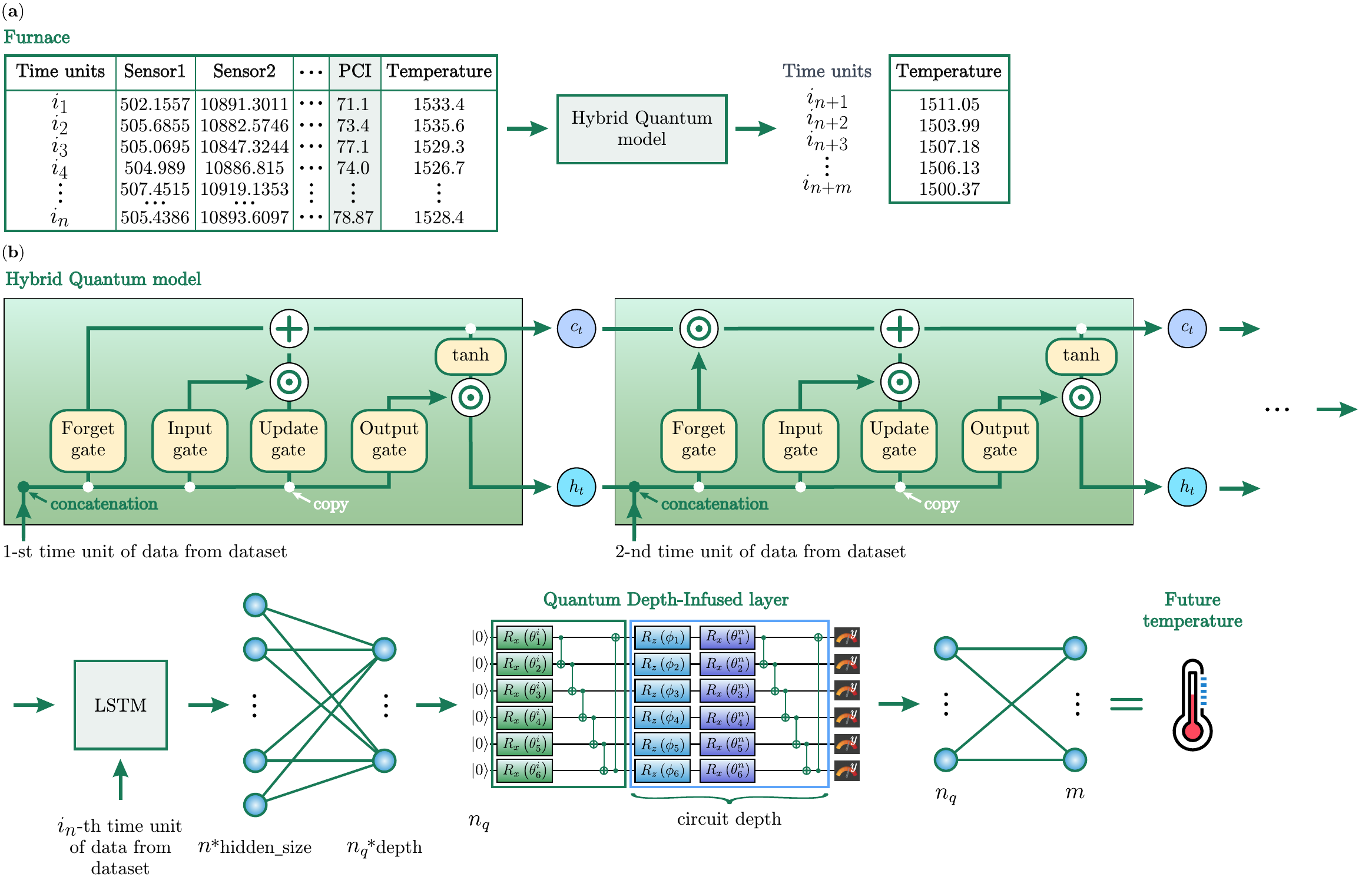}
    \caption{Architecture of the Hybrid Quantum Model for Temperature Prediction in a Blast Furnace: (a) The input consists of time-series sensor data from the blast furnace, including PCI. The model predicts future temperatures based on this data. (b) The core of the model is a Hybrid Quantum Neural Network combining a classical LSTM network with a QDI layer. The LSTM processes temporal sequences, then passes its output through a fully connected layer. The intermediate representation is encoded into quantum states using the QDI layer, which applies rotation and entanglement gates over 6 qubits. A final fully connected layer maps the quantum-encoded features to the predicted future temperature values.}
    \label{fig:hybrid}
\end{figure}

To further improve performance, we hybridized the above model by introducing a quantum layer \citep{Kordzanganeh2023}. The result is a hybrid quantum model for temperature prediction, shown in Fig.~\ref{fig:hybrid}. Its structure differs slightly from the classical model: we split the single fully connected layer into two layers, inserting a Quantum Depth-Infused (QDI) layer between them. The QDI layer acts as a feature transformer using a variational quantum circuit \citep{Sagingalieva2023c, Anoshin2024, Lusnig2024, Periyasamy2022, Sagingalieva2025}.

The QDI layer is designed to expressively encode a relatively large vector $n$ into the limited number of qubits $n_{\text{qubits}}$ available in the NISQ era. Its structure incorporates parameterized quantum rotation gates for angle encoding of data and trainable parameters, along with CNOT gates for entanglement between qubits. The rotation gates are defined as:
\begin{equation}
R_x(\theta) = \begin{pmatrix}
  \cos{\left(\frac{\theta}{2}\right)} & -i \sin{\left(\frac{\theta}{2}\right)} \\[4pt] 
  -i \sin{\left(\frac{\theta}{2}\right)} & \cos{\left(\frac{\theta}{2}\right)} 
\end{pmatrix}
\end{equation}

\begin{equation}
R_y(\theta) = \begin{pmatrix}
  \cos\left(\frac{\theta}{2}\right) & - \sin\left(\frac{\theta}{2}\right)\\[4pt] 
  \sin\left(\frac{\theta}{2}\right) & \ \cos\left(\frac{\theta}{2}\right)
\end{pmatrix}
\end{equation}

\begin{equation}
\text{CNOT} = \begin{pmatrix}
  1 & 0 & 0 & 0 \\
  0 & 1 & 0 & 0 \\
  0 & 0 & 0 & 1 \\
  0 & 0 & 1 & 0 \\
\end{pmatrix}
\end{equation}

In our implementation, we used $n_{\text{qubits}} = 6$ qubits. The QDI layer can be seen as a quantum neural network whose parameters are trained jointly with the classical parts of the model.

\subsection{Multi-feature Forecasting Model}

\begin{figure}[htbp]
    \includegraphics[width=\textwidth]{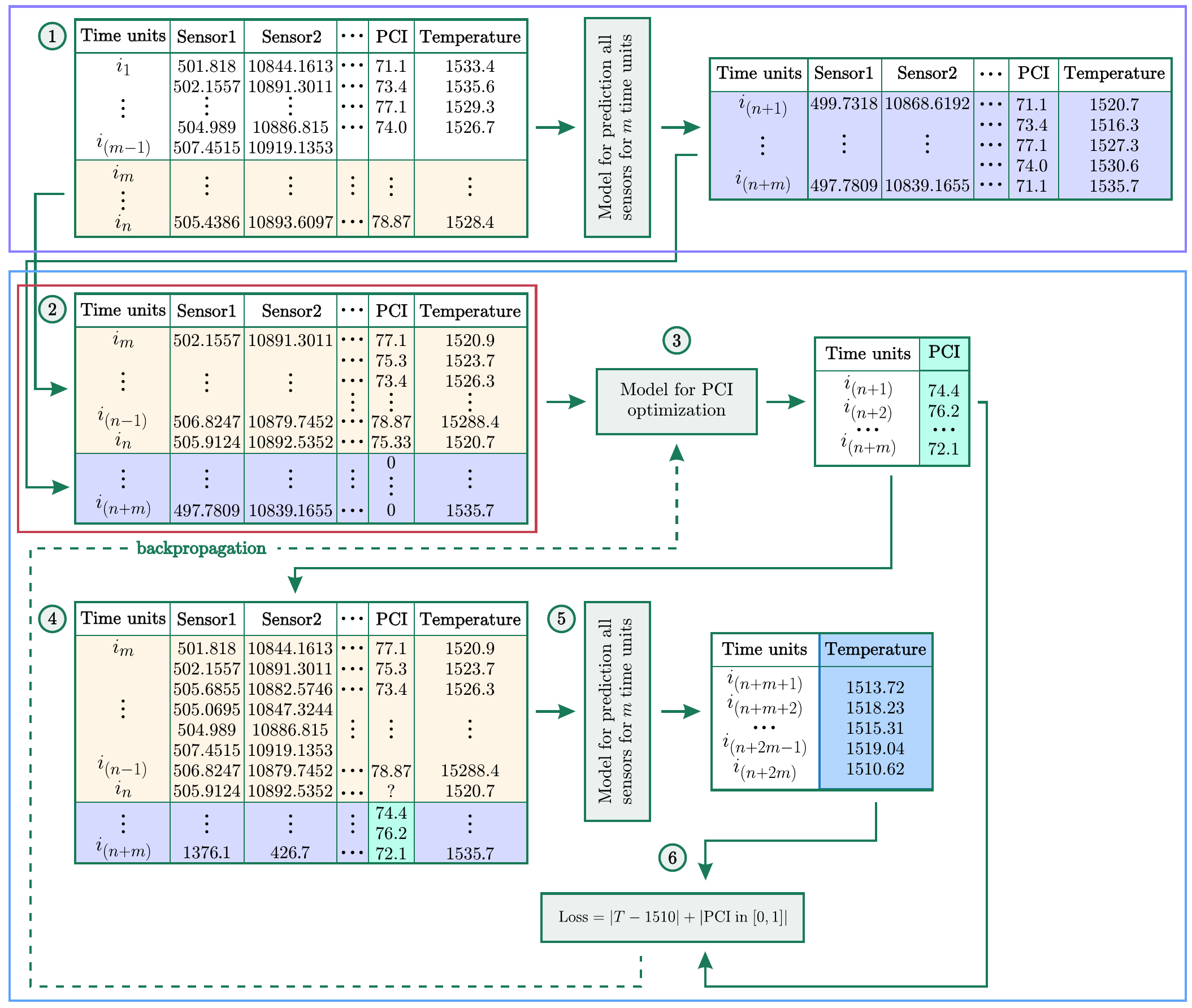}%
    \caption{Schematic representation of the optimization algorithm (Algorithm \ref{alg:opt}). Step 1: Predict baseline future: use model $\text{M}_\text{all}$ to predict all sensor features (including temperature) $m=5$ steps ahead from current state. Step 2: Augment with new PCI policy: concatenate predicted features with historical data and setting last $m$ PCI values to $0$. Step 3: Predict optimal PCI: use $\text{M}_\text{optim}$ to predict new PCI values for the next $m=5$ steps for balancing temperature (output shown in green box). Step 4 and 5: Simulate new policy outcome and predict the temperature: apply this proposed PCI policy (green box) in the $\text{M}_{\text{all}}$ model to predict the resulting temperatures (output shown in blue box). Compare the predicted temperatures to the target ($1510 \ ^\circ \text{C}$). Step 6: Update $\text{M}_\text{optim}$ (via backpropagation) to minimize the deviation (while also constraining PCI within $\left[0, 1\right]$). The dashed green line indicates the loss computation (difference between predicted temperature and target, plus a penalty for PCI bounds). Repeating this loop yields an optimized PCI policy that keeps the temperature within the target range.}
    \label{fig:Optimization}
\end{figure}
\label{opt}

Similar to the temperature-only model, we also constructed a model for predicting all 27 features ($\text{M}_\text{all}$) (including temperature and other key sensor readings) several steps into the future. The structure of $\text{M}_\text{all}$ is analogous to the single-output model: the LSTM layer uses a larger hidden state and the subsequent fully connected layers are wider, since the model now outputs a vector of 27 features rather than a single temperature for each of the 5 future time steps. Specifically, the input shape remains $24 \times 27$, but the output is $5 \times 27$ (predicting 5 future time steps for all 27 features).

\subsection{Models Evaluation}

We used two standard error metrics: Root Mean Square Error (RMSE) and Mean Absolute Error (MAE), defined as:
\begin{equation}
\text{RMSE}= \sqrt{\frac{1}{N}\sum^{N}_{t=1}(y^t-\hat{y}^t)^2}
\end{equation}

\begin{equation}
\text{MAE} = \frac{1}{N} \sum^{N}_{t=1} |y^t-\hat{y}^t|
\end{equation}

Our modeling approach is designed to run on both classical and quantum hardware using the same algorithm. At present, quantum processing units (QPUs) are still in early development \citep{Kordzanganeh2023b}, with most available quantum computers having fewer than 100 qubits and suffering from high error rates. Due to these limitations, we executed the hybrid model on classical hardware (simulating the quantum layer) for faster training and testing. However, as quantum hardware improves---especially with advances in error correction and qubit scalability---we expect significant gains in speed and computational efficiency for the quantum portion of the model.

\subsection{PCI Optimization Framework}

\begin{algorithm}[htbp]
\caption{Temperature-PCI Optimization}
\label{alg:opt}
\begin{algorithmic}
\Require Feature vectors $\{\ldots, \mathrm{PCI}, T\}_{t=t_i}^{t=t_i+23} \in \mathbb{R}^{24\times27}$; prediction model $\text{M}_{\text{all}}$; optimization model $\text{M}_{\text{optim}}$
\Ensure Optimized PCI policy $\widehat{\mathrm{PCI}}$ and corresponding temperature prediction $\widehat{T}$
\State \textbf{Step 1:} Map features to future predictions using $\text{M}_\text{all}$
\State \textbf{Step 2:} Stack truncated features with predicted values (PCI set to zero)
\State \textbf{Step 3:} Use $\text{M}_\text{optim}$ to predict optimization policy values
\State \textbf{Step 4--5:} Evaluate quality using $\text{M}_\text{all}$ to map new features to temperatures
\State \textbf{Step 6:} Tune $M_{\text{optim}}$ by minimizing composite loss:
$$ \mathcal{L} = \left\lVert \widehat{\{T\}} - 1510 \right\rVert_1 + \left\lVert \widehat{\{\mathrm{PCI}\}} \in [0, 1] \right\rVert_1$$
\end{algorithmic}
\end{algorithm}

Finally, we address the task of temperature optimization: choosing a sequence of PCI values (an injection policy) that will keep the furnace temperature within a desired range (e.g., $[1500\,^\circ\text{C}, 1510\,^\circ\text{C}]$). This is essentially a control problem where we use the predictive model to search for the optimal input sequence.

The optimization problem can be stated as: find a sequence of future PCI adjustments such that the predicted temperature stays in the target range. We propose a gradient-based search in the space of possible PCI sequences through an optimization model $\text{M}_\text{optim}$, which is implemented as an ordinary linear model.

The overall optimization procedure (summarized in Algorithm~\ref{alg:opt}) works as follows: use the multi-feature prediction model $\text{M}_{\text{all}}$ to simulate the effect of a candidate PCI policy on future temperatures, then iteratively adjust the PCI policy via gradient descent to minimize a cost function that penalizes deviation from the target temperature and impractical PCI changes.

\subsection{Training and Implementation Details}

The models $\text{M}_\text{T}$ and $\text{M}_\text{all}$ were trained on the entire historical dataset. We split the data into training and test sets with an 80:20 ratio. All input features were scaled using a MinMax scaler so that each feature falls between 0 and 1, which helps the neural networks train effectively.

Classical layer parameters were optimized via backpropagation \citep{Rumelhart1986}, implemented in PyTorch \citep{Paszke2019}, which computes loss gradients for gradient descent. The QDI layer in $\text{M}_\text{T}$ model requires specialized handling for quantum gradients, we used PennyLane \citep{Bergholm2018} with adjoint differentiation to optimize variational quantum parameters.

Table~\ref{tab:hyperparams} summarizes the key hyperparameters used in our experiments.

\begin{table}[htbp]
\centering
\caption{Hyperparameters used for model training.}
\label{tab:hyperparams}
\begin{tabular}{lc}
\toprule
Parameter & Value\\
\midrule
LSTM hidden size & 143\\
Number of qubits & 6\\
QDI circuit depth & 4\\
Input sequence length & 24 time steps\\
Output prediction horizon & 5 time steps\\
Number of features & 27\\
Train/test split & 80/20\\
Optimizer & Adam\\
Learning rate & 0.001\\
Batch size & 32\\
\bottomrule
\end{tabular}
\end{table}

\section{Results}
\label{sec:results}

\subsection{Feature Importance Analysis}

To analyze time-series correlations, a machine learning-based model called gradient boosting (GB) was used. This is an ensemble technique that combines multiple models to create a better model based on tree models. By performing GB, the influence of each variable on the final output can be calculated, which differs from simple distance-based correlation values.

The analysis identified the following highly correlated sensors: targeted PCI rate, hearth wall temperatures (8 sensors), elemental composition in SLAG (\ce{TiO_2}, \ce{MnO}, \ce{S}, \ce{SiO_2}), average charging time, STAVE temperatures (3 sensors), cooling plate temperatures (2 sensors), stack average temperature, real-time CR, BOSH gas volume, real-time FR, PCR, flame temperature, and OBYC.

From the results, a notable difference is observed between the values derived from influence calculation and those from simple correlation analysis, along with variations in rankings. Additionally, the influence-based time-series correlation analysis identified variables that were not previously well-monitored, such as SLAG \ce{TiO_2}, SLAG \ce{MnO}, SLAG S, and ObyC, thus making use of high-information-value data that had not been utilized effectively.

\subsection{Temperature Prediction Performance}

\begin{figure}[htbp]

    \includegraphics[width=\textwidth]{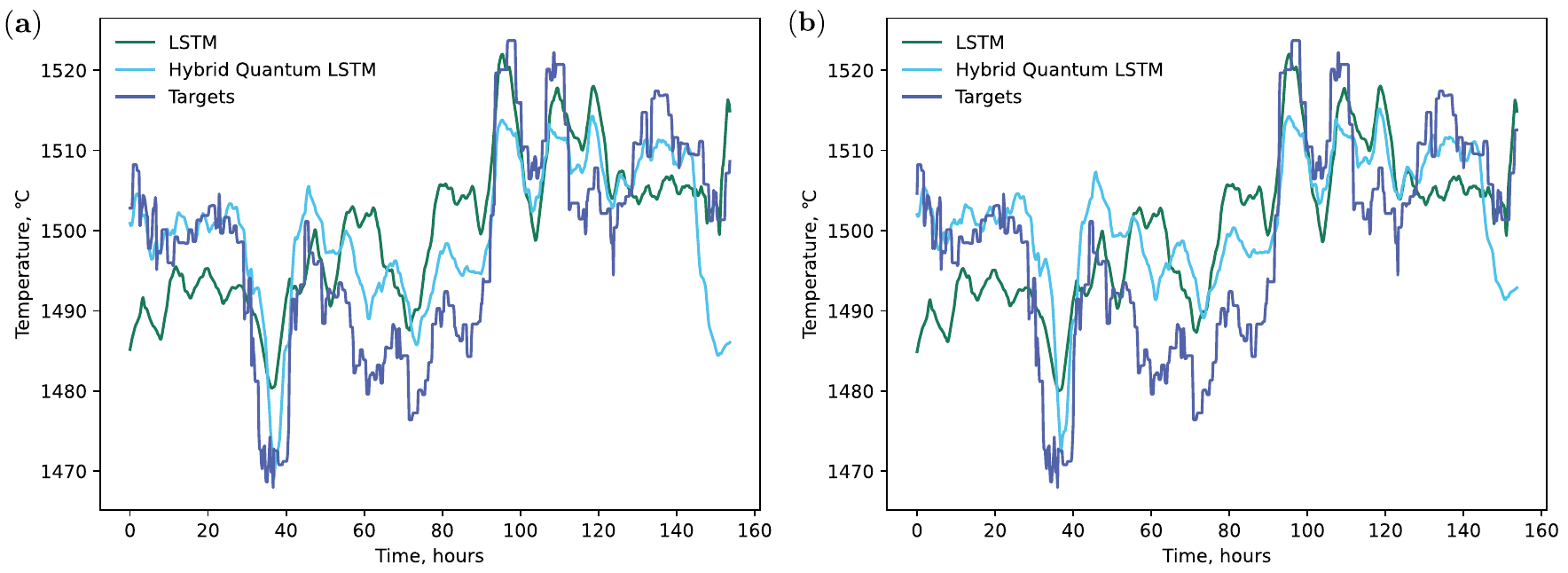}%
    \caption{Comparison of the temperature predictions between the LSTM and Hybrid Quantum LSTM models. (a) Predictions are made for 10 minutes forward (1 timestamp). Hybrid model outperforms the classical one with $\text{RMSE}_\text{q}=7.59$ and $\text{RMSE}_\text{cl}=9.98$. (b). Predictions are made for 50 minutes forward (5 timestamps). Hybrid model outperforms the classical one with $\text{RMSE}_\text{q}=8.55$ and $\text{RMSE}_\text{cl}=9.91$.}
    \label{fig:Prediction}
\end{figure}

In the furnace temperature prediction task, the hybrid quantum model demonstrated higher accuracy than the classical model. Table~\ref{tab:results} presents the comparative results for different prediction horizons.

\begin{table}[htbp]
\centering
\caption{Comparison of prediction accuracy between classical LSTM and hybrid quantum LSTM models.}
\label{tab:results}
\begin{tabular}{lccc}
\toprule
Model & Prediction Horizon & RMSE & MAE\\
\midrule
Classical LSTM & 10 min (1 step) & 9.98 & 7.82\\
Hybrid Quantum LSTM & 10 min (1 step) & 7.59 & 5.94\\
Classical LSTM & 50 min (5 steps) & 9.91 & 7.76\\
Hybrid Quantum LSTM & 50 min (5 steps) & 8.55 & 6.71\\
Classical LSTM & 1 hour & 5.95 & 4.66\\
Hybrid Quantum LSTM & 1 hour & 4.46 & 3.49\\
\bottomrule
\end{tabular}
\end{table}

For a 1-hour-ahead prediction, the hybrid model achieved an RMSE of 4.46 in normalized units corresponding roughly to $^\circ\text{C}$ on the test set, whereas the best classical model had an RMSE of 5.95. In percentage terms, the quantum-enhanced model showed up to a 25\% reduction in prediction error relative to the classical model without a quantum layer. This validates that even with current NISQ-era settings (simulated on classical hardware), adding a QDI layer can improve model performance, potentially by capturing complex feature interactions that the classical network struggled with.

Moreover, the hybrid model achieved this with fewer trainable parameters. This reduction in model size can be seen as a form of regularization, which may also contribute to the hybrid model's better generalization and lower overfitting tendency, consistent with observations in other HQNN studies \citep{Abbas2021, Berberich2024}.

\subsection{PCI Optimization Results}

Using the optimization framework described in Section~\ref{sec:methods}, we achieved temperature control within $\pm 7.6\,^\circ\mathrm{C}$ of the target temperature, compared to the previous variance of $\pm 50\,^\circ\mathrm{C}$. Figure~\ref{fig:optim} shows the comparison between the initial and optimized PCI policies and the resulting temperature trends.

\section{Discussion}
\label{sec:discussion}

\subsection{Practical Implications}

The achieved results have significant practical implications for steel manufacturing:

\begin{itemize}
\item \textbf{Energy Savings}: By reducing temperature variance from $\pm 50\,^\circ\mathrm{C}$ to $\pm 7.6\,^\circ\mathrm{C}$, operators can lower the target temperature setpoint, reducing fuel consumption. Given that approximately 110 tons/hr of coal is injected per blast furnace at costs of 30--60 thousand KRW per ton, even modest improvements translate to substantial cost savings.

\item \textbf{Process Stability}: Improved temperature control reduces the risk of dangerous events such as ``hang-ups'' and ``back lamination,'' enhancing worker safety and equipment longevity.

\item \textbf{Quality Consistency}: Stable temperatures lead to more consistent molten iron quality, reducing downstream production disruptions.
\end{itemize}

\subsection{Computational Considerations}

With sufficient iterative computation, our optimization approach was able to derive PCI guidance values that reduce post-control temperature fluctuations to within about $\pm 7.6\,^\circ\mathrm{C}$ of the target. However, to deploy this in real time on an actual blast furnace, we must consider the system latency and how many iterations can feasibly be run between measurement updates. In practice, the number of optimization iterations would be limited by computation time and the frequency of new data.

An important direction for future work is to test how the performance degrades when limiting the number of algorithm iterations or the prediction horizon. The goal will be to find a balance between optimization thoroughness and real-time responsiveness.

\subsection{Limitations and Future Work}

Several limitations and opportunities for future work should be noted:

\begin{enumerate}
\item \textbf{Quantum Hardware}: Currently, simulations were used due to limitations of available quantum hardware. 

\item \textbf{Multi-feature Quantum Model}: We did not implement a quantum layer in the multi-output model ($\text{M}_\text{all}$) due to computational resource requirements. A hybrid quantum $\text{M}_\text{all}$ could be developed as quantum simulators improve.

\item \textbf{Generalizability}: While our work focused on blast furnace temperature control, the framework is general and could be applied to other industrial control problems.
\end{enumerate}

\section{Conclusion}
\label{sec:conclusion}

In this work, we introduced a complete machine learning pipeline for predictive blast furnace control, consisting of sophisticated data preprocessing, feature selection, and a suite of predictive models, including hybrid quantum-enhanced models to describe the internal furnace processes. Building on this foundation, we proposed a machine-learning-driven optimization approach for PCI injection rate control.

The developed models, including the hybrid quantum model incorporating a QDI layer, can predict the furnace temperature and other sensor data with high accuracy. The hybrid quantum temperature model achieved an RMSE of 4.46 for a 1-hour ahead prediction, compared to the purely classical model---roughly a 25\% improvement in accuracy.

These predictive models lay the groundwork for accurate temperature control, which we implemented via an optimizer model. Using a linear regression optimizer embedded in the pipeline, the furnace temperature was controlled within roughly half of the previously required range, achieving an accuracy of about $7.6\,^\circ\mathrm{C}$ around the target temperature. This level of stability successfully meets the stringent requirements of steel production for hearth temperature control.

Overall, the achieved results, both in terms of prediction accuracy and PCI policy optimization, exceed what is typically seen in contemporary industrial practice. This demonstrates the viability and advantage of applying QML models to real-world industrial problems. The success of our approach suggests that even early quantum technologies, when thoughtfully integrated into classical frameworks, can provide tangible benefits in complex industrial scenarios. We believe this work is a significant step toward the practical adoption of QML in industry, and we anticipate that continued improvements in quantum hardware will further amplify these benefits.

\section*{Declarations}

\subsection*{Funding}
No funding was received for conducting this study.

\subsection*{Competing Interests}
The authors declare no competing interests.

\subsection*{Author Contributions}
Nayoung Lee and Minsoo Shin contributed to data curation, industrial domain expertise, and validation. Asel Sagingalieva contributed to methodology and quantum computing framework development. Arsenii Senokosov contributed to software implementation and data analysis. Matvei Anoshin contributed to model development. Ayush Joshi Tripathi and Karan Pinto contributed to experimental design and analysis. Alexey Melnikov supervised the project and contributed to methodology. All authors reviewed and approved the final manuscript.

\subsection*{Data Availability}
The datasets generated and/or analyzed during the current study are proprietary to POSCO and are not publicly available due to confidentiality agreements. Requests for data access may be directed to POSCO Holdings Inc.

\subsection*{Ethics Approval}
Not applicable. This study does not involve human participants or animals.

\subsection*{Consent to Participate}
Not applicable.

\subsection*{Consent for Publication}
Not applicable.

\section*{Acknowledgements}
The authors thank POSCO for providing access to industrial data and domain expertise.

\end{document}